\documentstyle[seceq,mbf,epsf,wrapft]{ptptex}

\title{
Quasi-Particle Spectrum around a Single Vortex
in Superconductors
}
\subtitle{s-Wave Case}    

\author{
Masaru {\sc Kato}\footnote{E-mail: kato@ms.osakafu-u.ac.jp}
and Kazumi {\sc Maki}$^{*,}$\footnote{E-mail: kmaki@usc.edu}
}

\inst{
Department of Mathematical Sciences, Osaka Prefecture University\\
Sakai Osaka 599-8531, Japan
\\
$^*$Department of Physics and Astronomy,
University of Southern California\\
Los Angeles, CA 90089-0484, USA}

\recdate{December 20, 1999
}

\abst{
Making use of the Bogoliubov-de Gennes equation,
we study the quasi-particle spectrum and the vortex core structure
of a single vortex in quasi 2D $s$-wave superconductors
for small $p_F\xi_0$, where $p_F$ is the Fermi momentum and
$\xi_0=v_F/\Delta_0$ is the coherence length ($\hbar=1$).
During our numerical calculation the particle number is conserved for
each $p_F\xi_0$.
In particular, we find that there are only 1 or 2 bound states for
$p_F\xi_0=1$.
 Also, for $p_F\xi_0=1$,
the Kramer-Pesch effect ceases to exist at  around $T/T_c\simeq 0.3$.
}

\begin{document}

\maketitle

\section{Introduction}
There are renewed interests in vortex structure
since the discovery of high $T_c$ cuprate superconductors.
It is possible that the superconductivity of high $T_c$ cuprates
can be characterized
as $d$-wave superconductivity\cite{tsuei,harlinger}
and is very close to the quantum limit.\cite{maki,doett}

Schopohl and Maki previously studied the quasi-particle spectrum
around a single vortex line in terms of a quasi-classical equation
\cite{eilen,larkin} and predicted a clear four-fold symmetry
for a $d$-wave superconductor.\cite{schopohl}
Then, a beautiful STM study of vortices in YBCO monocrystals
was reported.\cite{maggio}
There are several interesting results of these studies.
First, it was found that the quasi-particle spectrum exhibits circular
symmetry; there is no trace of four-fold symmetry
Second, there appears to be only a single bound state
with energy
$\approx \frac{1}{4}\Delta(0)$, where $\Delta(0)$ ($=260 \mbox{ K}$)
is the superconducting order parameter at $T=0 \mbox{ K}$.
A similar bound state energy was observed earlier by a far-infrared
magneto-transmission from a YBCO film by Karra\"{\i} et al.\cite{karrai}
According to the analysis of bound states around the vortex core by Caroli
and co-workers,~\cite{caroli}
there should be thousands of bound states.
Of course, in usual $s$-wave superconductors
we have $p_F\xi_0\approx 10^3\sim10^4$, where $p_F$ is the Fermi momentum and
$\xi_0=v_F/\Delta(0)$ is the coherence length ($\hbar=1$).
Thus perhaps the single bound state in YBCO suggests
$p_F\xi_0\approx 1$.\cite{maki}
At first sight this suggestion seems outrageous, since this implies
$E_F\approx 200\sim 500 \mbox{ K}$ in YBCO.
By analyzing the spin gap at $T=0$ K observed in YBCO monocrystals with
inelastic neutron scattering by Rossat-Mignod et al.,\cite{rossat}
we can deduce chemical potential $\mu$ given as\cite{won}
\begin{equation}
\mu = -345(x-0.45) \mbox{ K},
\end{equation}
where $x$ is the oxygen dopage corresponding to YBa$_2$Cu$_3$O$_{6+x}$.
Then, for optimally doped YBCO we obtain $\mu =-190 \mbox{ K}$.

In D\"{o}ttinger et al.,\cite{doett} the flux-flow resistance of 60 K
YBCO measured by Matsuda et al.\cite{matsuda} is analyzed,
and they identified the Kramer-Pesch effect characteristic of a
superconductor in the clean limit.\cite{K-P,L-O}
On the other hand, apparently the Kramer-Pesch effect is absent
in 90 K YBCO,\cite{matsuda} implying again thar perhaps 90 K YBCO
is in the quantum limit.\cite{doett}
Here, the Kramer-Pesch effect causes the vortex core to shrink
with decreasing temperature due to the decrease in the occupied
bound states around the vortex line.
The core size can be expressed as\cite{K-P}
\begin{equation}
\xi_1=\frac{v_F}{\Delta(T)}\frac{T}{T_c}.
\end{equation}
This reduction of the core size results
in nonlinear conductivity.\cite{doett,L-O}
Ichioka et al.\cite{ichi} found the Kramer-Pesch effect
in $d$-wave superconductivity with the help of a semi-classical approach
developed in Ref.~\citen{schopohl}.

But if $p_F\xi_0\approx 1$, this implies that the semi-classical
approach introduced in Refs.~\citen{eilen} and \citen{larkin}
is no longer reliable in studying the vortex in high $T_c$ cuprates.
For this reason, Morita, Kohmoto and Maki\cite{morita} studied the
Bogoliubov-de Gennes equation for a single vortex in $d$-wave superconductors.
Indeed, choosing $p_F\xi_0=1.33$, they were able to describe
gross features of the STM result for YBCO.
On the other hand, they have not attempted the
self-consistent calculation due to numerical difficulties.
Nevertheless, they discovered that there is
a single bound state for $p_F\xi_0=1.33$
in the vortex of $d$-wave superconductors.
Further, there are low energy ($E\leq0.1\Delta$) extended states
with four legs stretched in the four diagonal directions $(\pm1,\pm1,0)$.
Recently, these extended states were rediscovered by Franz and Te\v{s}anovi\'{c}
\cite{franz} in a model somewhat different from
that in Ref.~\citen{morita}.
Further, Franz and Te\v{s}anovi\'{c} claimed there should be no bound states,
based on a different model which contains at least 3 arbitrary parameters.
We believe that the absence of bound states is either due to a) neglect of
particle number conservation or b) the rather strong Coulomb repulsion
they introduced.
It is rather surprising that Yasui and Kita\cite{yasui} and Takigawa et al.\cite{takigawa}
obtained similar results as Ref.~\citen{franz}.
Clearly in these later works, particle number conservation has been neglected.
On the other hand, we have shown recently 
there  are a few bound states within the weak
coupling model, as in Ref.~\citen{morita}, down to $p_F\xi_0\simeq1$.\cite{kato}
Also, this shows clearly that number conservation is crucial, as pointed out
by van der Marel.\cite{marel}
Also, the above results confirm the validity of the work in Ref.~\citen{morita}
rather than that in Ref.~\citen{franz}.

In this paper we study the quasi-particle spectrum around a single vortex in $s$-wave
superconductors in the quantum limit.
This problem was previously considered by Hayashi et al.\cite{hayasi}
However, as we have mentioned, they ignored the number conservation,
which is very crucial.
For example van der Marel has shown that the chemical potential depends
on the temperature if number conservation is imposed.~\cite{marel}
Further, in Ref.~\citen{hayasi}, the energy cutoff is made arbitrarily, so that
their result is not reliable, as we will show.
Also, as we will show, the $r$-dependence of $\Delta(r)$ near $r=0$
is completely determined by $u_0(r)v_0(r)$,
where $u_0(r)$ and $v_0(r)$ are spinor wave functions for the lowest bound state.

In the following we concentrate on the cases $p_F\xi_0=1,2\mbox{ and }4$
for simplicity, and study the quasi-particle spectrum and the shape of
$\left|\Delta\left(r\right)\right|$ as a function of temperature.
A preliminary result on this has appeared in Ref.~\citen{kato2}.

\section{Bogoliubov-de Gennes equation}
The spatial dependence of order parameters in superconductivity
is described by the Bogoliubov-de Gennes equation.
We consider the case at a small magnetic field near $H_{c1}$, so that
there is a single vortex.
Therefore we can ignore the vector potential.
In this case, the Bogoliubov-de Gennes equation becomes
\begin{subeqnarray}\label{BGORIG}
\left(-\frac{1}{2m_e}\nabla^2-\mu \right)u_n\left({\mbf r}\right)+
\Delta\left({\mbf r}\right)v_n\left({\mbf r}\right)
=E_nu_n\left({\mbf r}\right), \\
-\left(-\frac{1}{2m_e}\nabla^2-\mu \right)v_n\left({\mbf r}\right)+
\Delta^*\left({\mbf r}\right) u_n\left({\mbf r}\right)
=E_n v_n\left({\mbf r}\right),
\end{subeqnarray}
where $u_n\left({\mbf r}\right)$ and $v_n\left({\mbf r}\right)$ are
quasi-particle wave functions.

We take the $z$-axis to be along the vortex line.
We consider the nearly two-dimensional case for simplicity,
where the kinetic term associated with the z direction is negligible.
In the following, we merely consider the two-dimensional case, and
we use cylindrical coordinates.
Taking the gauge as $\Delta({\mbf r})=|\Delta(r)|e^{-i\theta}$,
the angular momentum of each
eigenstate becomes half of an odd integer, $m+\frac{1}{2}$.\cite{caroli}
Then $u_n({\mbf r})$ and $v_n({\mbf r})$ become as follows;
\begin{subeqnarray}
u_n(r,\theta)=u_{n m}(r)\frac{e^{im\theta}}{\sqrt{2\pi}},\\
v_n(r,\theta)=v_{n m}(r)\frac{e^{i(m+1)\theta}}{\sqrt{2\pi}}.
\end{subeqnarray}
Following Gygi and Schl\"{u}ter,\cite{gygi-schl}
we apply the Fourier-Bessel expansion with basis
\begin{equation}
\phi_{mj}(r)=\frac{\sqrt{2} }{RJ_{m+1}(\alpha_{jm})}
J_m(\alpha_{jm}\frac{r}{R})
\end{equation}
to $u_{n m}(r)$ and $v_{n m}(r)$, where $\alpha_{jm}$ is the $j$-th positive zero of
the Bessel function of $m$-th order, $J_m(x)$.
Then the wave functions become as
\begin{subeqnarray}
u_{n m}(r)=\sum_ju_{n m j} \phi_{m j}(r),\\
v_{n m}(r)=\sum_jv_{n m j} \phi_{m+1 j}(r).
\end{subeqnarray}
Here, the boundary condition is such that  wave functions are zero
at the edge of the disk with radius $R$.
Then the Bogoliubov-de Gennes equation becomes
\begin{subeqnarray}
\left[\frac{1}{2m_e}\left(\frac{\alpha_{jm}}{R}\right)^2
-\mu\right]u_{n m j}+\sum_{j_1}\Delta_{jj_1}v_{n m j_1}
=E_{n m}u_{n m j},\\
-\left[\frac{1}{2m_e}\left(\frac{\alpha_{j\ m+1}}{R}\right)^2
-\mu\right]v_{n m j}+\sum_{j_1}\Delta_{j_1j}u_{n m j_1}
=E_{n m}v_{n m j},\label{eq:BdG}
\end{subeqnarray}
where $\Delta_{j_1j_2}$ is given as
\begin{equation}
\Delta_{j_1j_2}=\int_0^R\phi_{m j_1}(r)\left|\Delta(r)\right|
\phi_{m+1j_2}(r)r dr.
\end{equation}
The order parameter is given as
\begin{equation}
\left|\Delta(r)\right|=
g\sum_{\left|E_{nm}\right|\leq E_c}\sum_{m\ge0}\sum_{j_1j_2}
u_{nm j_1}v_{nm j_2}\phi_{m j_1}(r)
\phi_{m+1j_2}(r)[1-2f(E_{nm})],\label{eq:OP}
\end{equation}
where $g$ is the interaction constant, $E_c$ is the cutoff energy, and $f(E)$ is
the Fermi distribution function.

Also we impose particle number conservation,
\begin{subeqnarray}
 N&=&2\int \sum_n \left\{ \left| u_n \left( {\mbf r} \right) \right|^2
 f\left( E_n \right)+\left| v_n \right|^2
 \left(1-f \left(E_n \right) \right) \right\} d^2{\mbf r},\\
 &=&2\sum_j\sum_{nm}\left[\left|u_{nmj}\right|^2f\left( E_{nm} \right)
 +[\left|v_{nmj}\right|^2\left(1-f \left(E_n \right) \right) \right],\label{eq:ncon}
\end{subeqnarray}
where $N$ is the total number of particles.
The chemical potential is determined by this equation.
Here $N$ is the particle number of the normal state with specified $p_F$ at $T=0$.
We solve these equations, Eqs.~(\ref{eq:BdG}), (\ref{eq:OP})
and (\ref{eq:ncon}),
self-consistently.
We fix the cutoff energy ($E_c=5 \Delta_0$
for the case of largest Fermi momentum)
and the interaction constant, and choose the
Fermi momentum $p_F$ so that $p_F\xi_0=1, 2\  \mbox{and}\  4$.

\section{Results}

In the numerical calculation, the radius of boundary is taken as $R=10\xi_0$,
and maximum angular momentum and number of zero points
of the Bessel functions are
taken so that all of the quasi-particle states within the cutoff energy
are taken into the calculation.

\begin{figure}
\epsfxsize=13.5cm
\centerline{\epsfbox{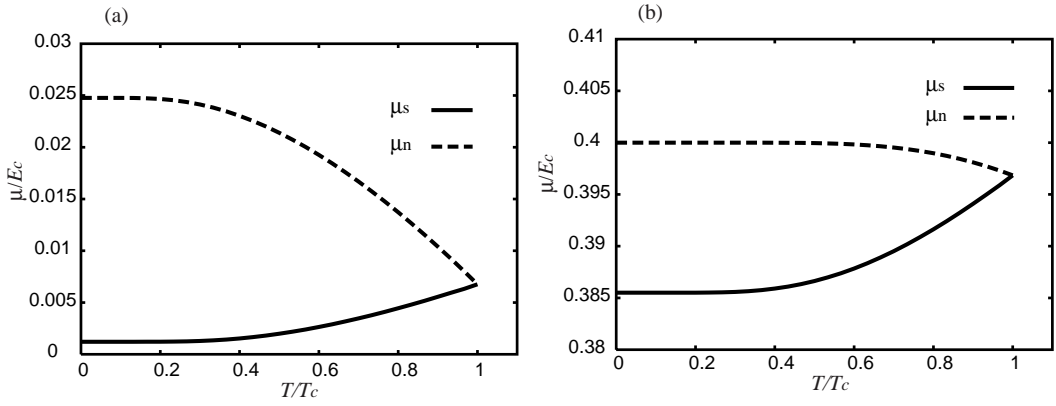}}
\caption{Temperature dependence of the chemical potential for
$p_F\xi_0=1$ (a) and $p_F\xi_0=4$ (b).
The solid curve is the chemical potential for the superconducting state,
and the broken curve is that for the normal state.}
 \label{MU}
 \end{figure}

Before proceeding to the numerical results, 
here we discuss the temperature dependence of the chemical potential
for the superconducting state. 
The chemical potentials for the normal state ($\mu_n$) and
the superconducting state ($\mu_s$) of the two-dimensional system are
given as\cite{marel}
\begin{eqnarray}
\beta\mu_n\left(T\right)&=&\ln\left(e^{\beta\mu_n\left(0\right)}-1\right),
\\
2\mu_n\left(0\right)&=&\mu_s\left(T\right)
+E_c-\sqrt{E_c^2+\Delta\left(T\right)^2}+
\sqrt{\mu_s\left(T\right)^2+\Delta\left(T\right)^2}\nonumber\\
&&-\frac{2}{\beta}\ln\frac{1+e^{-\beta\sqrt{E_c^2+\Delta\left(T\right)^2}}}
{1+e^{-\beta\sqrt{\mu_s\left(T\right)^2+
\Delta\left(T\right)^2}}}+\frac{2}{\beta}\ln\left(1+e^{-\beta E_c}\right).
\end{eqnarray}
The temperature dependences of the chemical potential for
$p_F\xi=1 \mbox{ and } 4$ are shown in Fig.~\ref{MU}.
As pointed out by van der Marel,\cite{marel} the chemical potential
decreases with decreasing temperature in the superconducting state.
This is opposite to the behavior in the normal state.
As we can see from the figure for $p_F\xi_0=1$,
the chemical potential becomes very small in the 
superconducting state, and the chemical potential of the normal state
drops significantly from $T=0$ to $T=T_c$.
For larger $p_F$, this temperature dependence of the chemical potential
becomes much weaker.
Therefore the temperature dependence and the $p_F$ dependence
of physical quantities may change from those with
the constant chemical potential calculation in the quantum limit.

\subsection{Temperature dependence of order parameter}
\label{sec:op}
\begin{figure}[htb]
\parbox{\halftext}{
\epsfxsize=6cm
\epsfbox{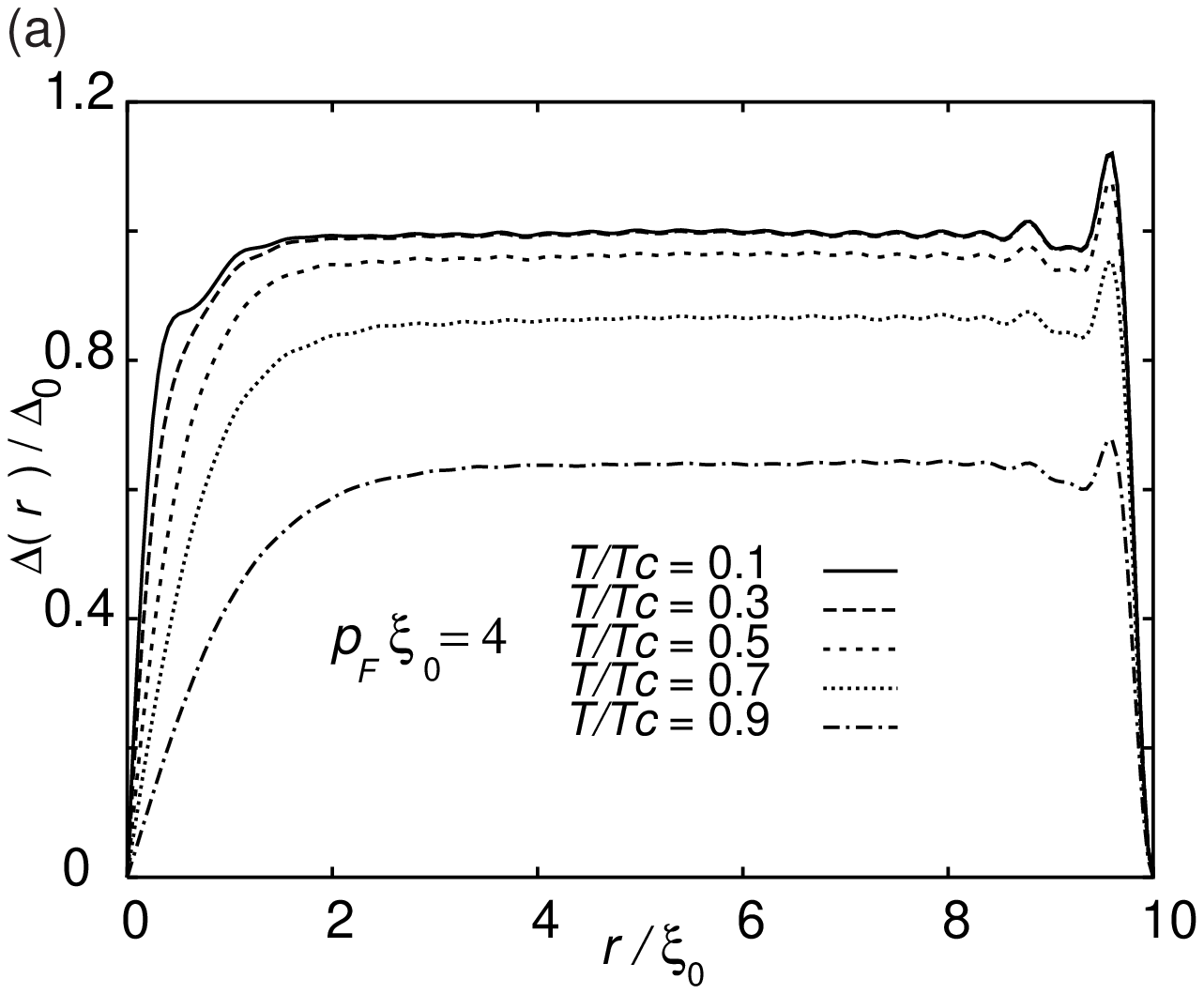}
\epsfxsize=6cm
\epsfbox{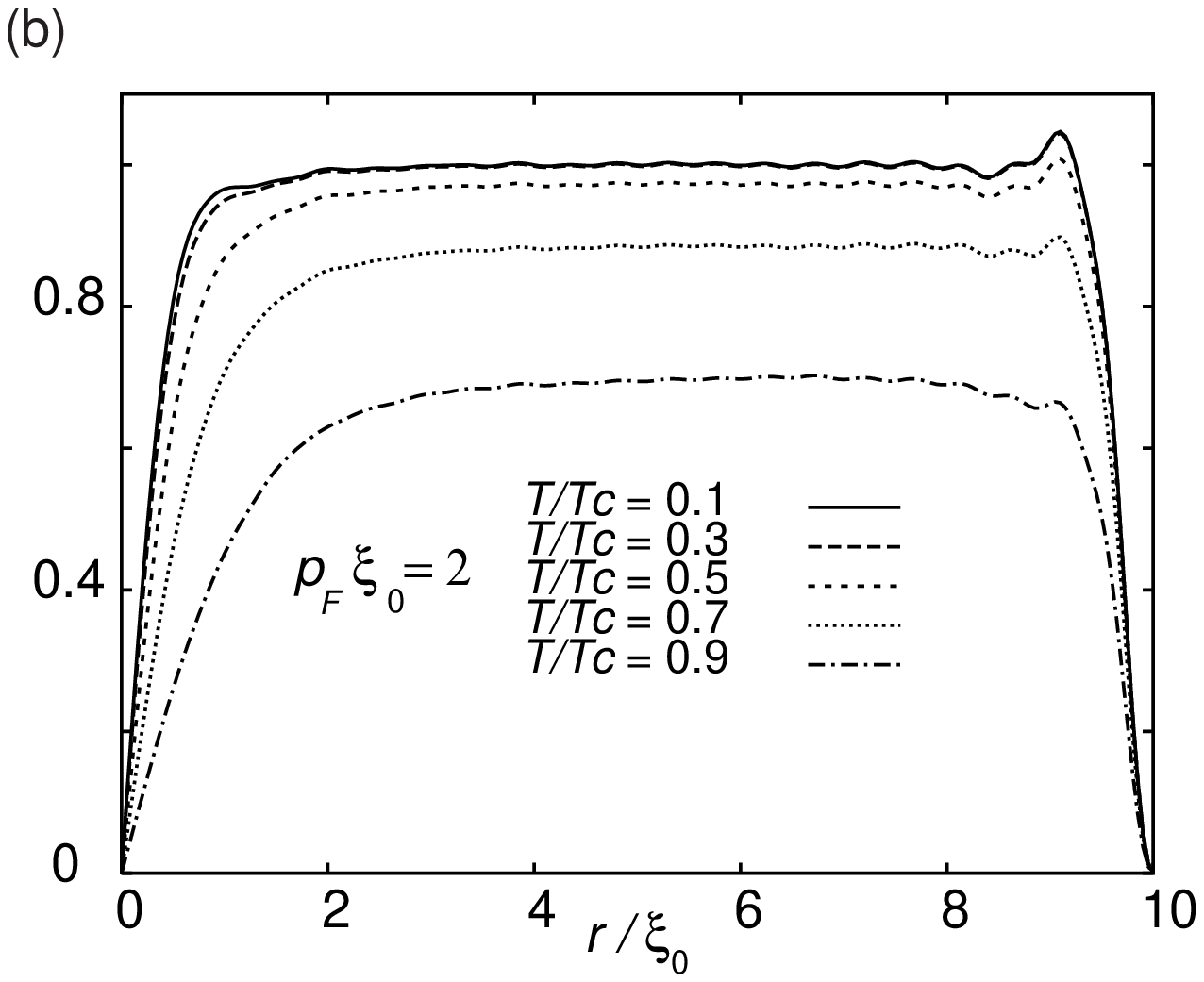}
\epsfxsize=6cm
\epsfbox{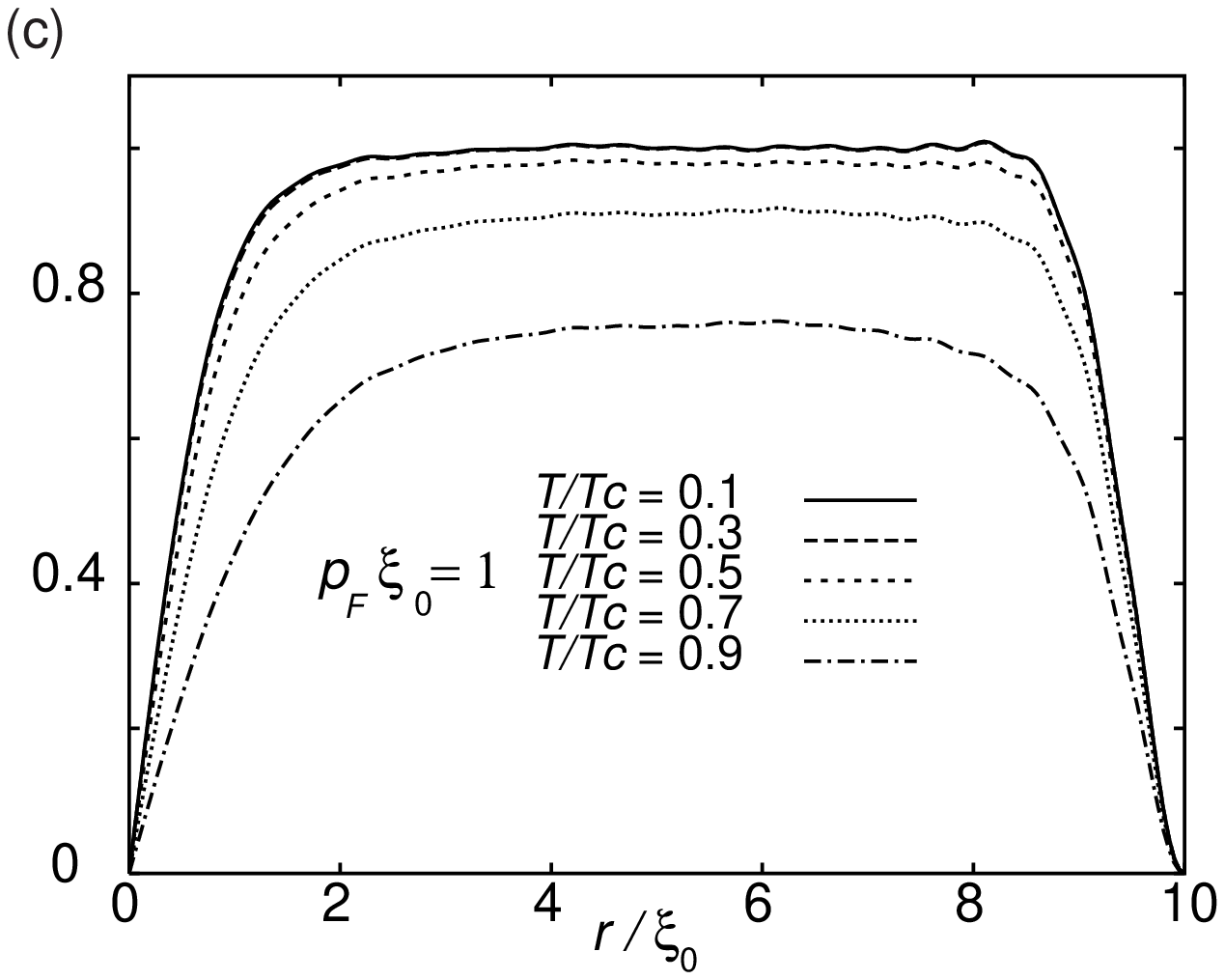}
\caption{The temperature dependence of the order parameter $\Delta(r)$
for (a) $p_F\xi_0=4$, (b) $p_F\xi_0=2$ and (c) $p_F\xi_0=1$.
The order parameter is normalized with $\Delta_0=\Delta(r=5.5\xi_0, T=0)$
for each $p_F\xi_0$.}
\label{OP}}
\hspace{8mm}
\parbox{\halftext}{
\epsfxsize=6.6cm
\epsfbox{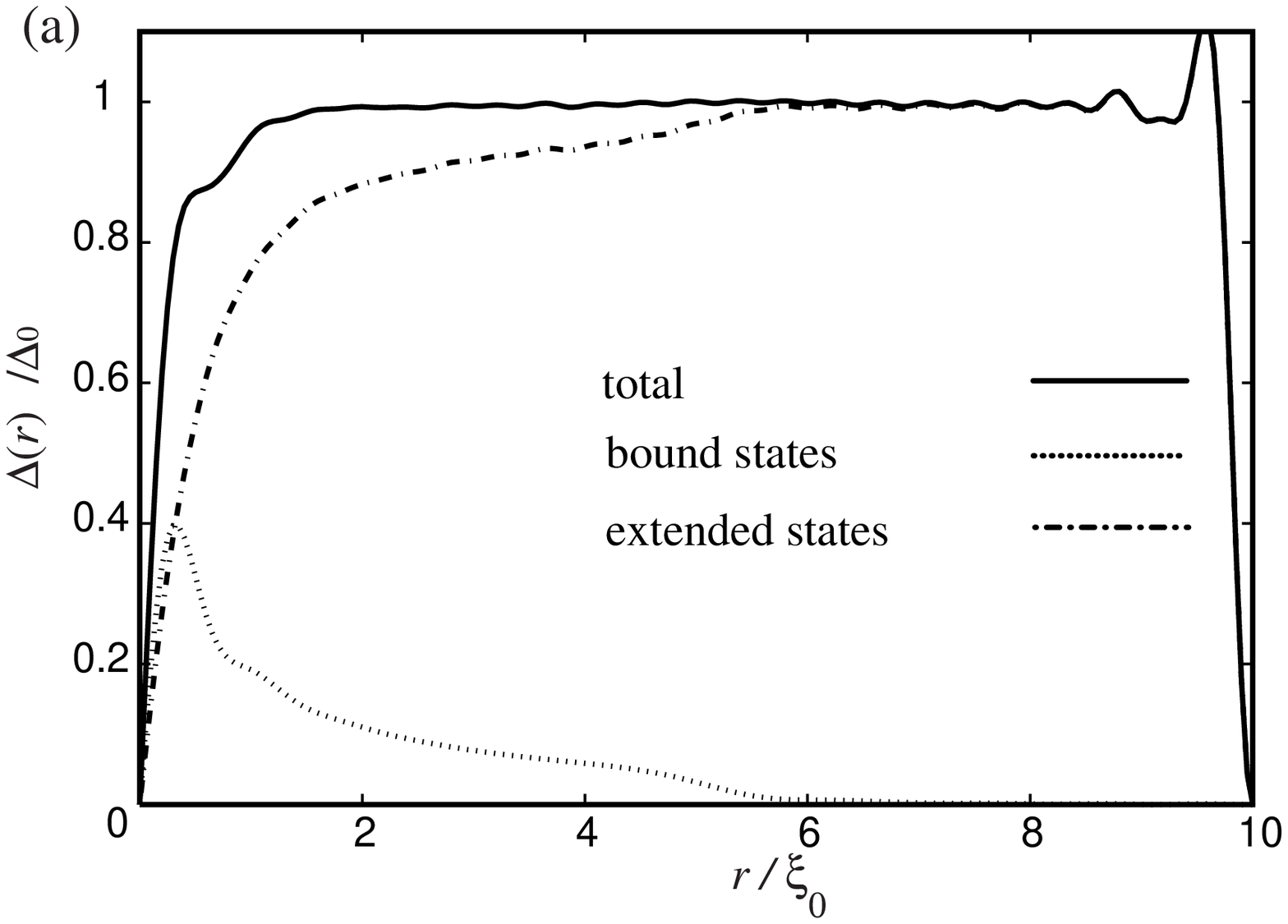}
\epsfxsize=6.6cm
\epsfbox{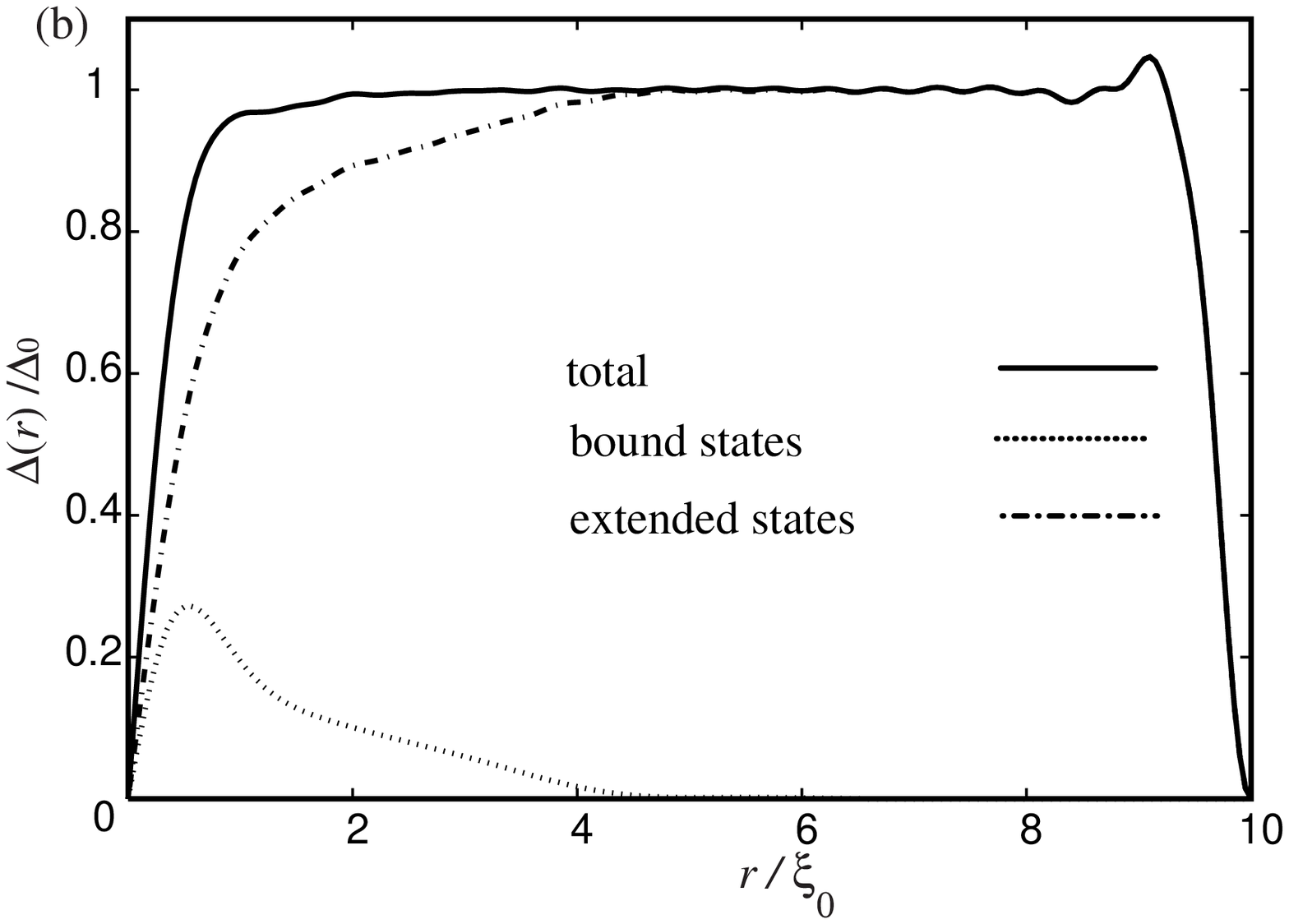}
\epsfxsize=6.6cm
\epsfbox{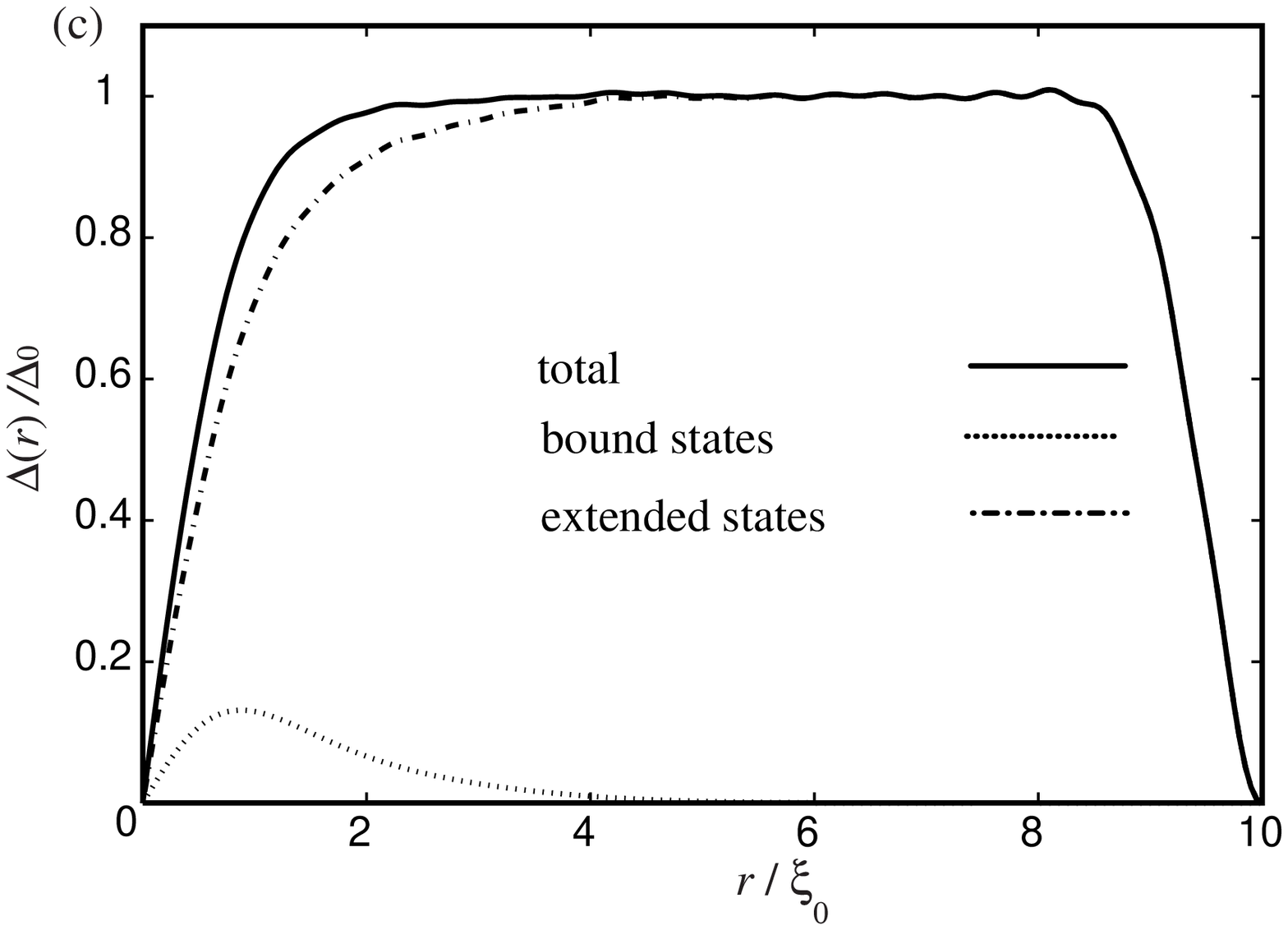}
\caption{Contributions from the bound states and the scattering states
to the order parameter for (a) $p_F\xi_0=4$, (b) $p_F\xi_0=2$
and (c) $p_F\xi_0=1$, where $\Delta_0=\Delta(r=5.5\xi_0, T=0)$.}
\label{OPsep}}
\end{figure}

The temperature dependence of the order parameter is shown in Fig.~\ref{OP}.
Near the boundary $r/\xi_0=10$,
the order parameter goes to zero, and there is a peak before that.
These are effects from the boundary condition and the finite size.
But the core structures are not affected by the boundary condition.

For $p_F\xi_0=4$ at low temperature ($T/T_c\leq0.1$)
there is a shoulder in $|\Delta(r)|$ at the vortex core.
This feature comes from bound states in the vortex core.
This can be seen from Fig.~\ref{OPsep},
where contributions of the scattering states and the bound states are shown
separately.

For  $p_F\xi_0=1$, the peak position of
the contribution to the order parameter is located slightly outside
of the vortex core.
Therefore the core structure is not so strongly affected by the bound states.
For larger $p_F\xi_0(\sim 16)$, Hayashi et al.\cite{hayasi}
demonstrated the oscillation
of the order parameter on the inside of the vortex core.
They argued that this oscillation is Friedel oscillation,
but from our figures it appears that the origin of the oscillation is the discreteness 
of the bound states.

Slightly increasing the temperature,
the bound state contribution decreases rapidly,
and the scattering state contribution remains almost the same.
In this temperature range, the core size is dominated by the bound states.
Above this temperature region, the scattering state contribution decreases with
increasing temperature, which is the same behavior
as that of the order parameter of the uniform solution.

\begin{figure}[htb]
 \parbox{\halftext}{
 \epsfxsize=6.6cm
\epsfbox{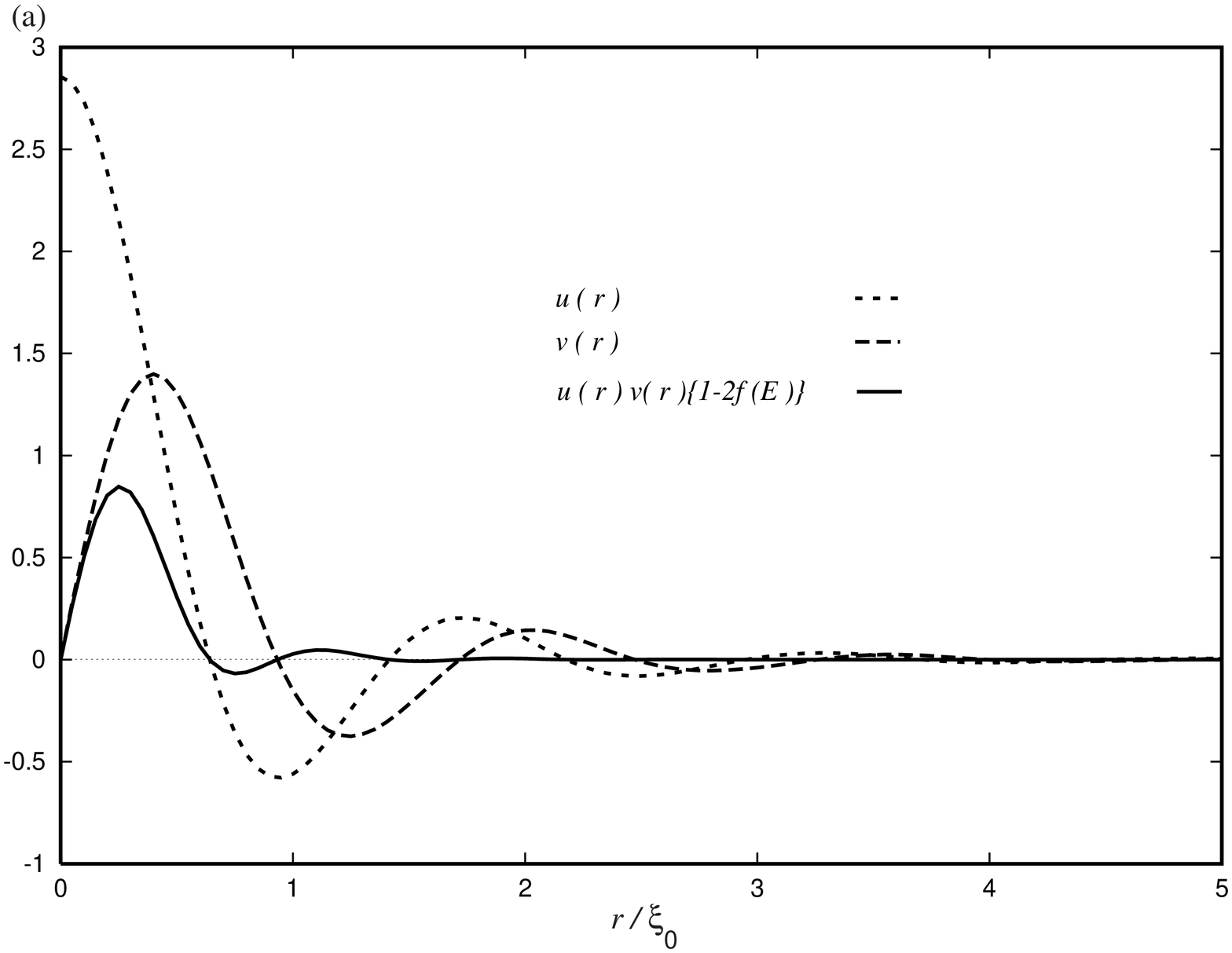}
  \epsfxsize=6.6cm
\epsfbox{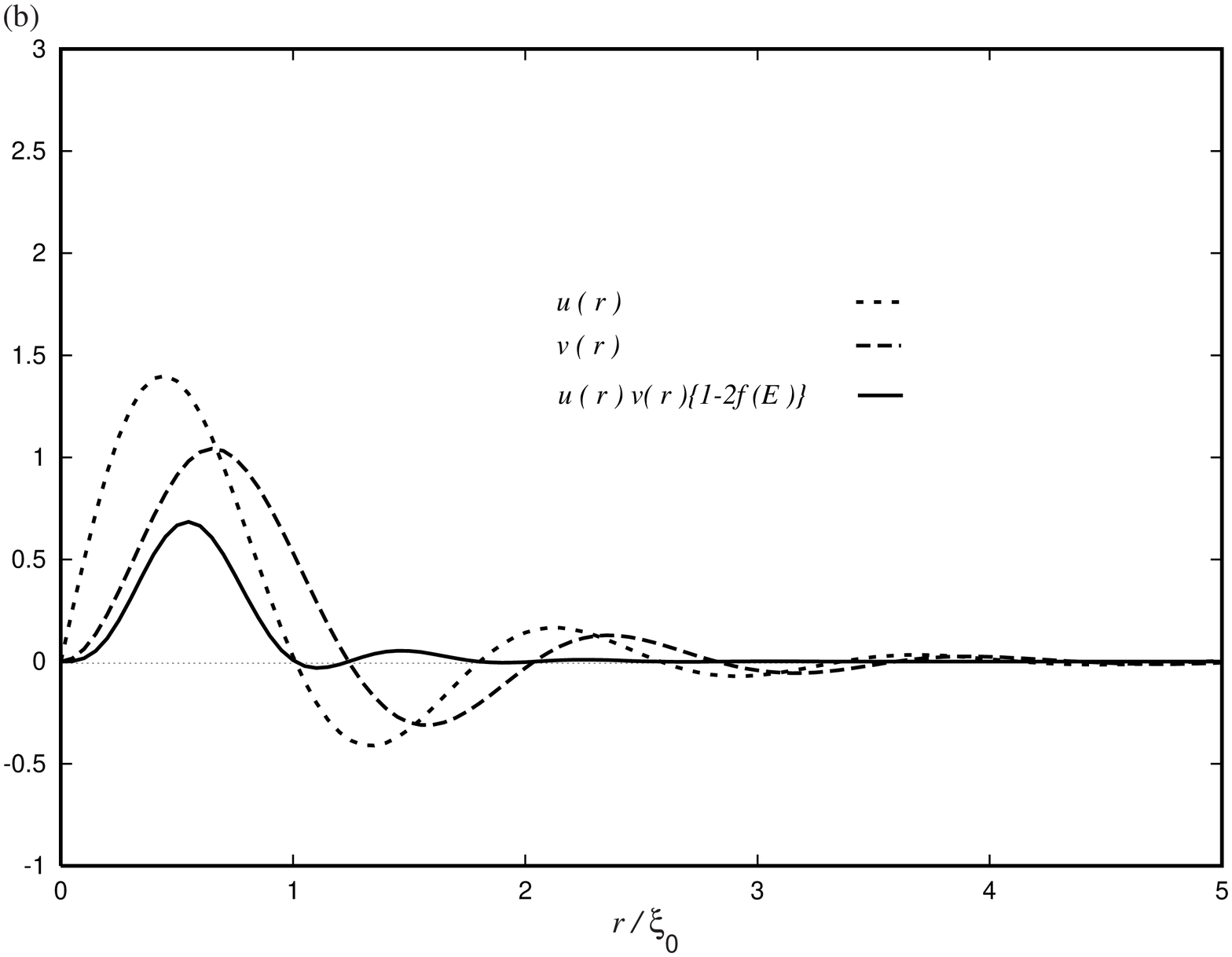}
  \epsfxsize=6.6cm
\epsfbox{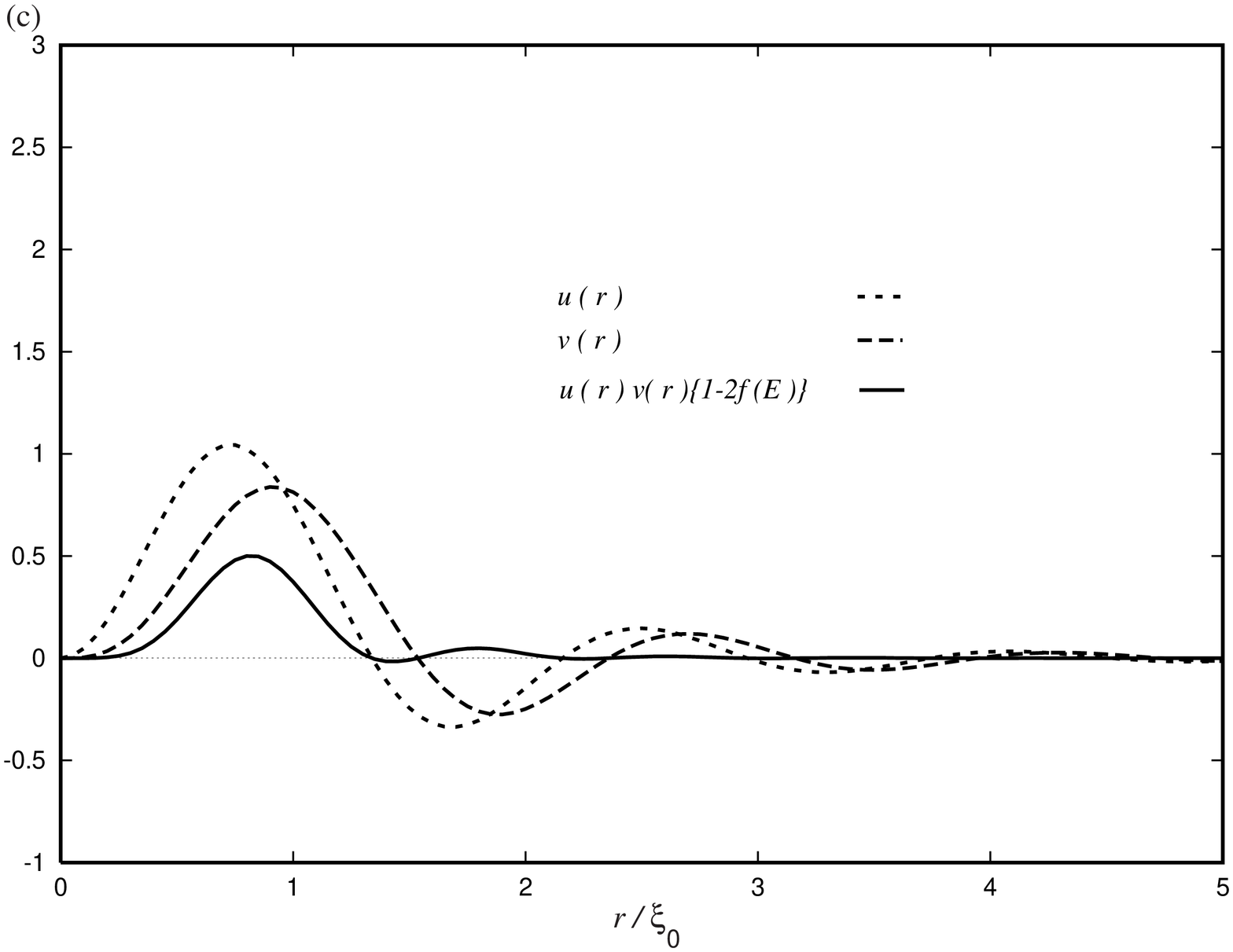}
  \caption{The quasi-particle wave functions $u(r)$ and $v(r)$
and their product
with a thermal factor for (a) the first, (b) the second and (c) the third
of the lowest energy bound states.}
 \label{amp}}
 \hspace{8mm}
\parbox{\halftext}{
\epsfxsize=6.6cm
\epsfbox{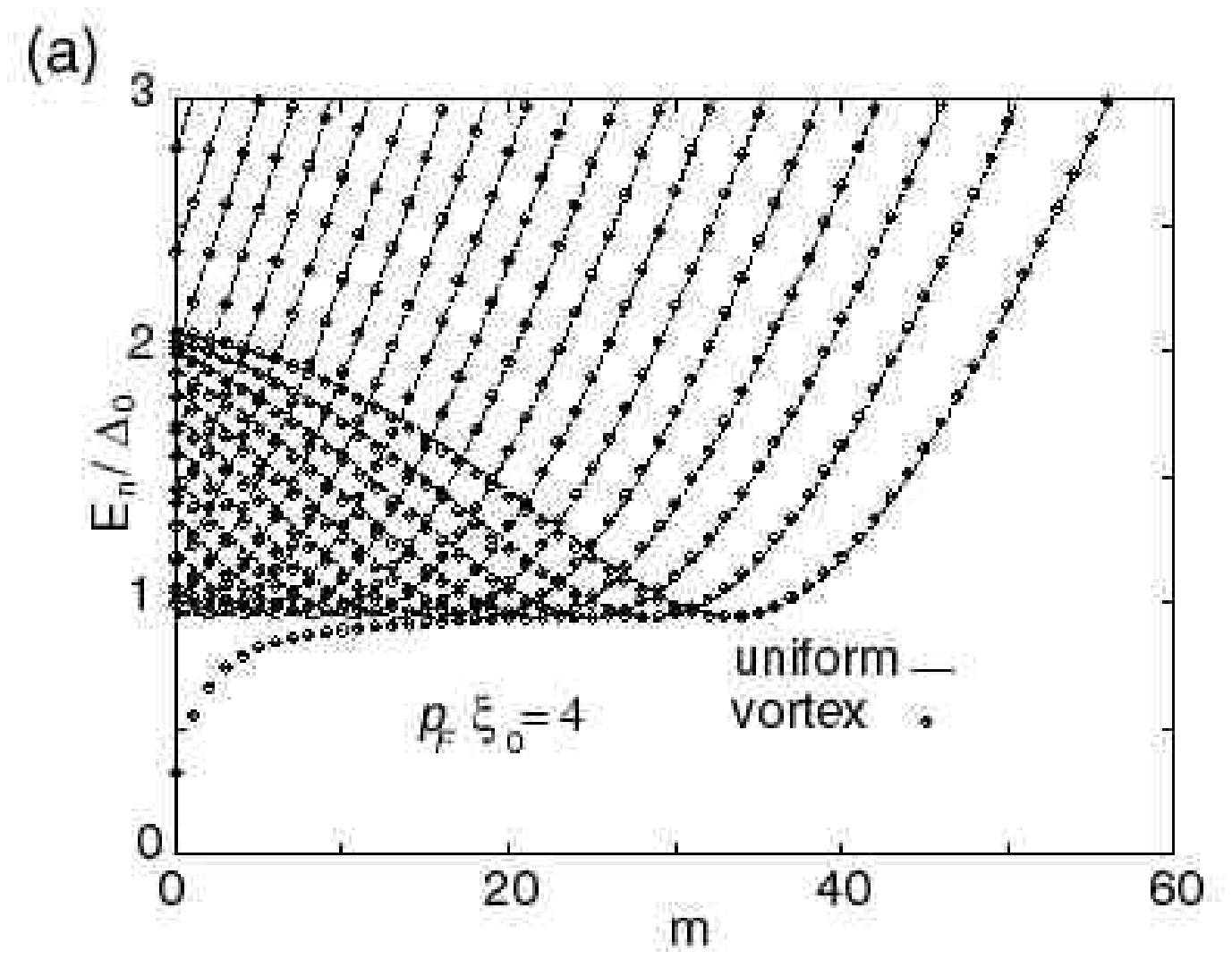}
  \epsfxsize=6.6cm
\epsfbox{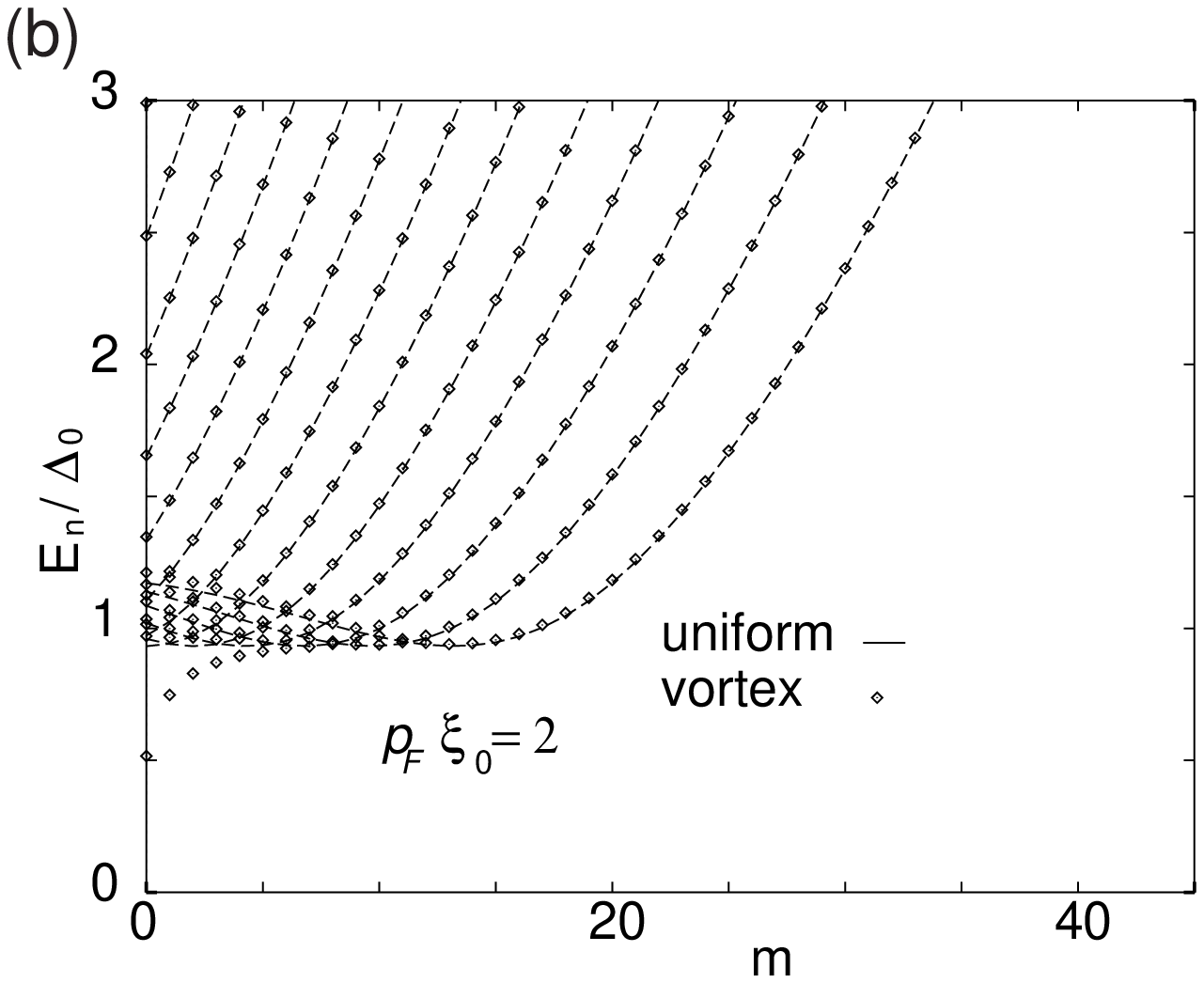}
  \epsfxsize=6.6cm
\epsfbox{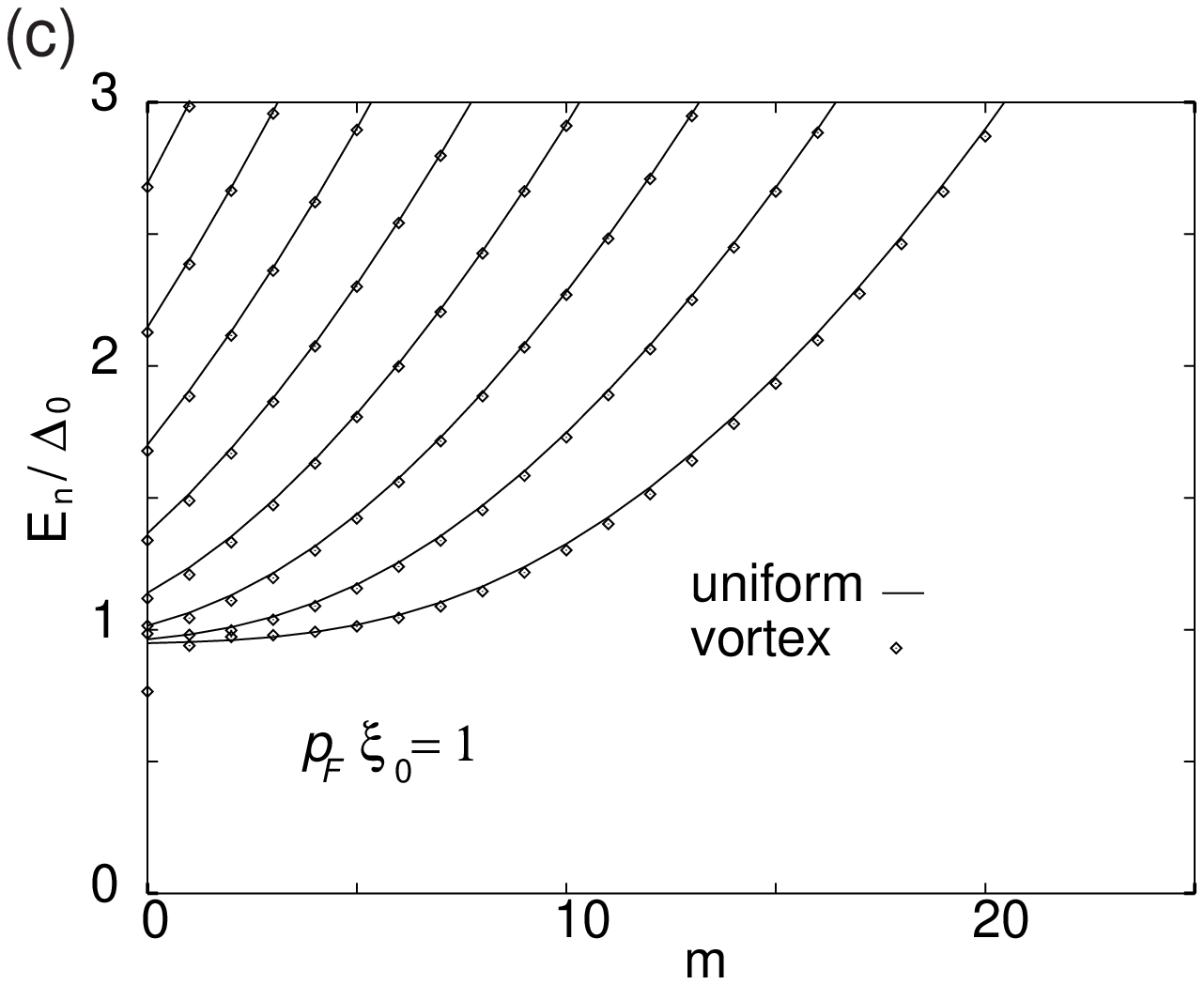}
   \caption{The quasi-particle spectrum of the vortex state ($\Diamond$)
and the uniform state (dashed curves)
for (a) $p_F\xi_0=4$, (b) $p_F\xi_0=2$ and (c) $p_F\xi_0=1$.}
 \label{excit}}
\end{figure}

We also plot the quasi-particle wave functions $u_{n\mu}(r)$ and $v_{n\mu}(r)$
of three  bound states for $p_F\xi_0=4$ in Fig.~\ref{amp}.
From these figures, it can be seen that
main contribution to the core structure comes from the lowest energy
bound state.

Also, we can see that $u(r)$ of the lowest energy bound state
behaves like an $s$-wave function, and $v(r)$ behaves like $p$-wave function.
Similar behavior also can
be seen for the second and third lowest bound states.
They belong to the angular momentum $m$ and $m+1$ states, respectively.

\subsection{Quasi-particle spectrum}
In Fig.~\ref{excit} we give quasi-particle energies with
angular momentum $m+\frac{1}{2}$, and we also plot those of the uniform state
in order to compare the vortex state and the uniform state.
For small angular momentum, the lowest energy states become bound states,
lowering their energies and slightly increasing the energies of  higher energy states.
For larger angular momentum, the quasi-particle energy is almost same
for vortex and uniform states, and there is no bound state.
\begin{wrapfigure}{l}{6.6cm}
 \epsfxsize=6cm
  \centerline{\epsfbox{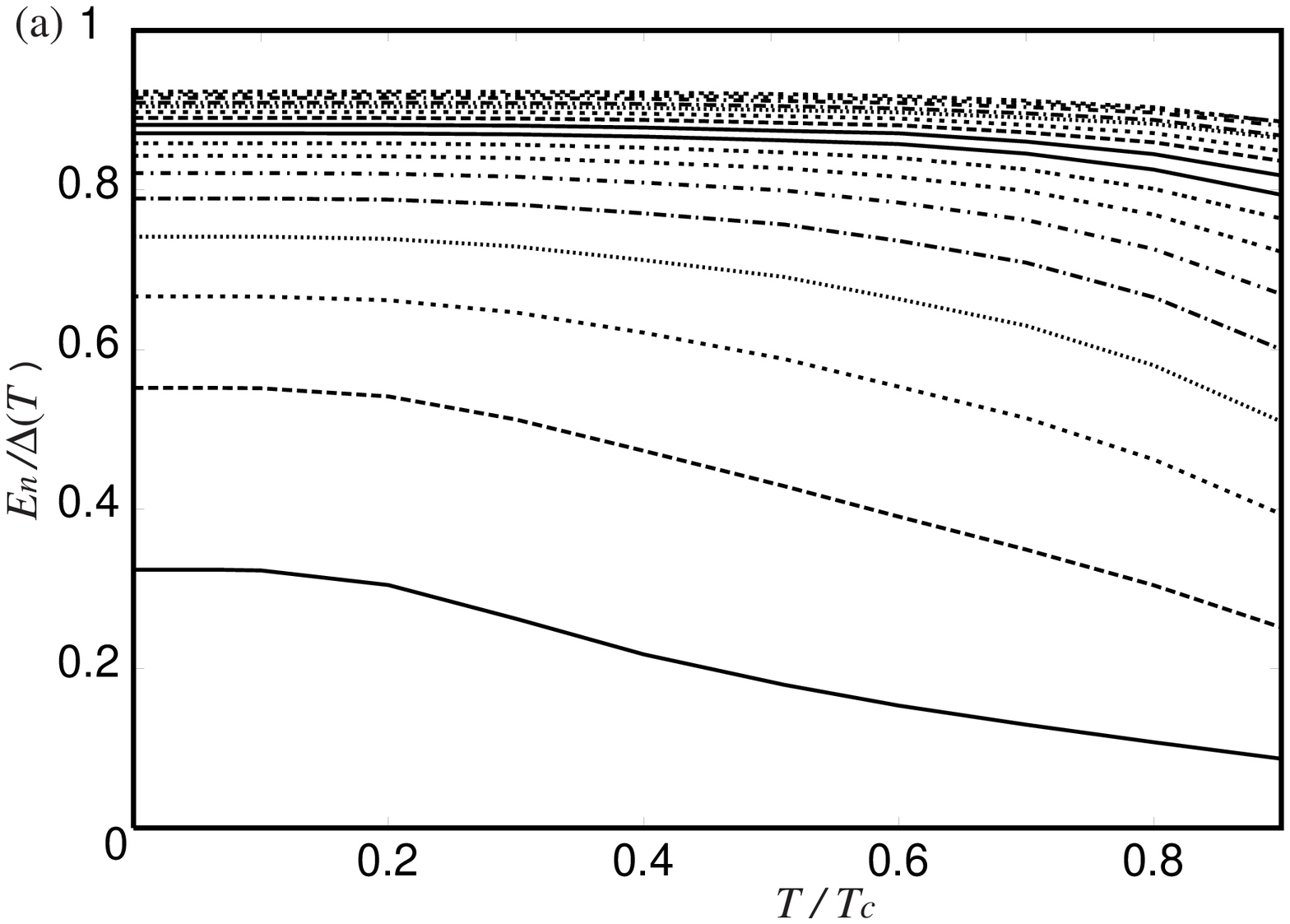}}
  \epsfxsize=6cm
  \centerline{\epsfbox{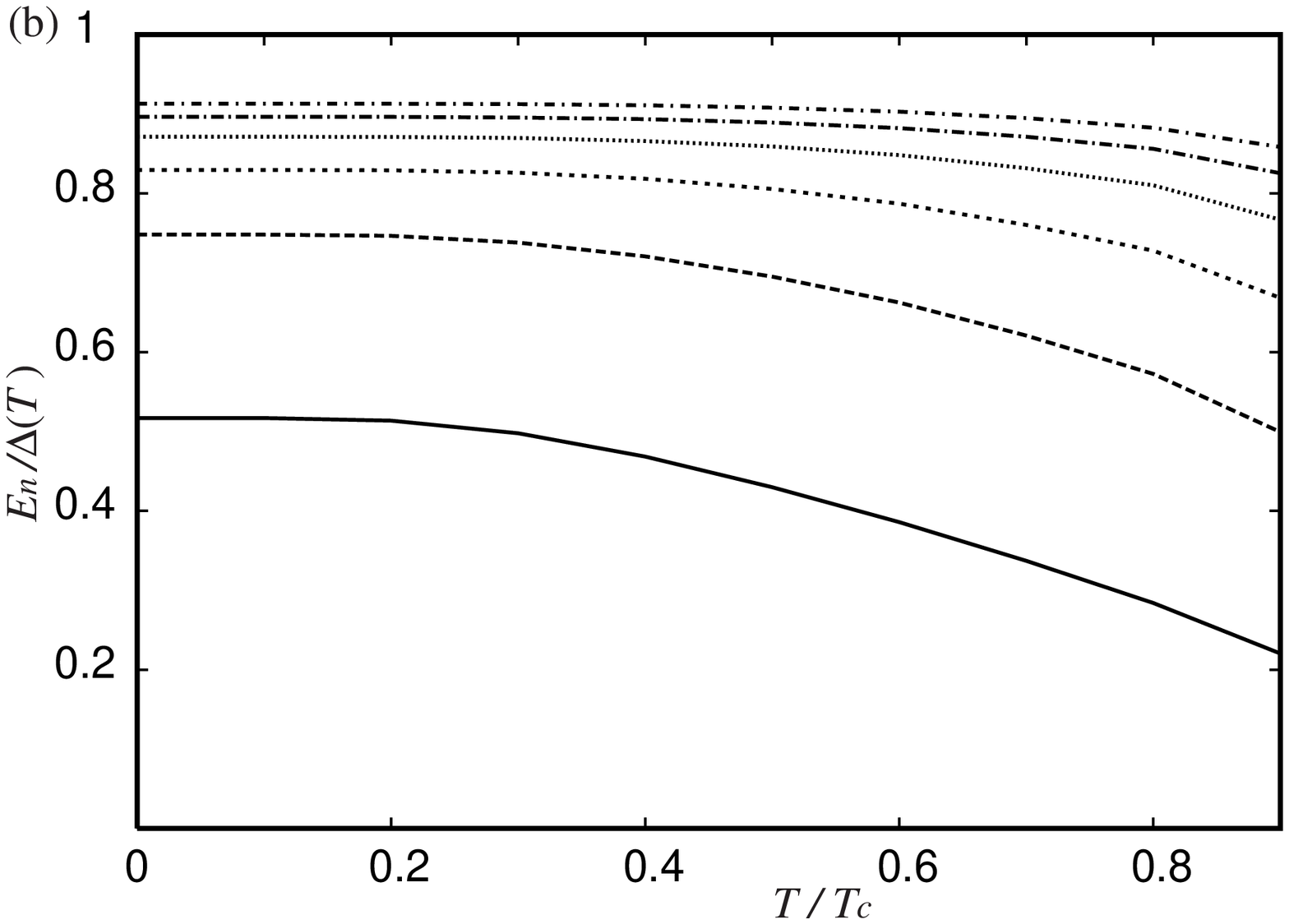}}
  \epsfxsize=6cm
  \centerline{\epsfbox{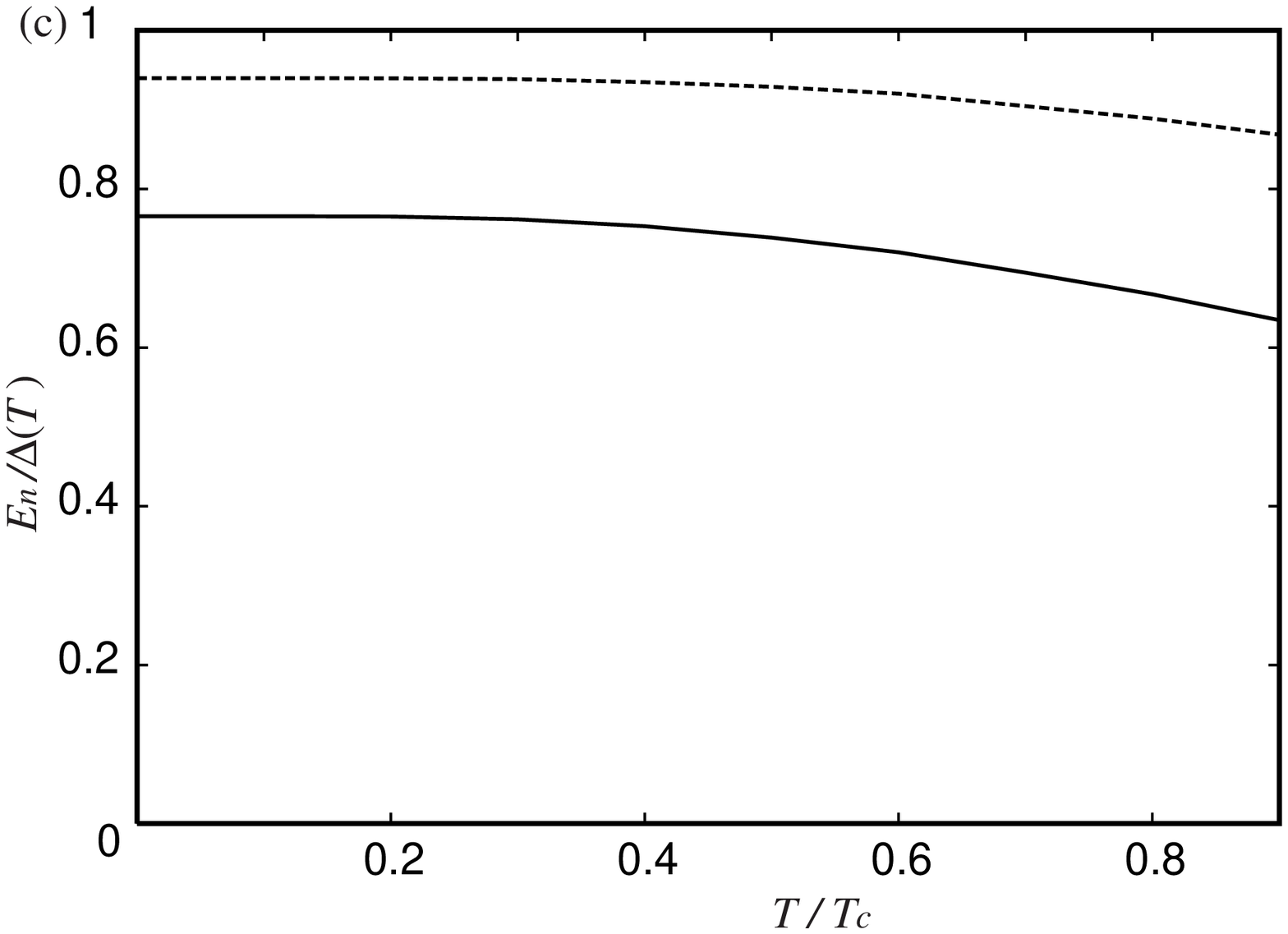}}
\caption{The temperature dependence of the quasi-particle energies
of the bound states in the vortex state.
The energies are normalized by the order parameter $\Delta(T)$
at $r/\xi_0=5.5$. (a), (b) and (c) are for $p_F\xi_0=4$, $p_F\xi_0=2$
and  $p_F\xi_0=1$, respectively.}
 \label{excit-ft}
\end{wrapfigure}
Although it is difficult to determine the boundary of these two cases
in the angular momentum, the number of bound states may be counted
as $18$ for $p_F\xi_0=4$, 5 for $p_F\xi_0=2$, and 1 or 2 for $p_F\xi_0=1$.
Compared with our previous calculation without number conservation,
the number of the bound states decreases. This effect is clearer for
smaller $p_F\xi_0$.
 
Such a relation between the number of the bound states and Fermi momentum 
is important for $d$-wave superconductors, because the local density of states
changes greatly due to this relation.
Therefore, it is worthwhile to point out this relation for the $s$-wave case, and
in order to obtain the correct relation, our number conserved calculation is 
needed.

The temperature dependence of the quasi-particle energy
is shown in Fig.~\ref{excit-ft}.
 In this figure we plot the quasi-particle energies of several bound states
normalized with the order parameter $\Delta(T)$ at $r=5.5\xi_0$.
From this figure we can see that for smaller $p_F\xi_0$ and
higher energy states, the quasi-particle energy varies in proportion to
the order parameter with temperature.
But for lower energy states with larger $p_F\xi_0$, the quasi-particle energy
increases rapidly with decreasing temperature and becomes constant
at low temperature.
This behavior of the energy of the lowest bound state
was previously noted by Gygi and Schl\"{u}ter\cite{gygi-schl} for much larger
$p_F\xi_o$,
but their result shows that the energy increases with decreasing temperature
even at low temperature because of the large value of $p_F\xi_0$.

\subsection{Core radius}
\begin{wrapfigure}{r}{6.6cm}
  \epsfxsize=6.6cm
  \centerline{\epsfbox{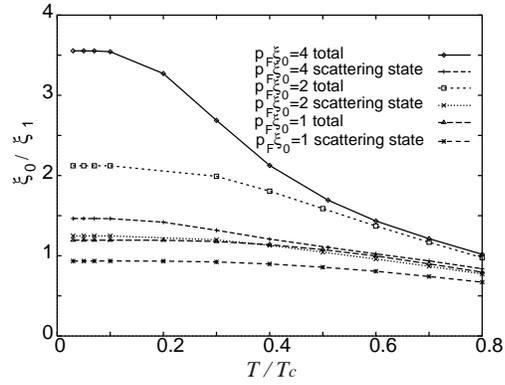}}
 \caption{The temperature dependence of
$\xi_1^{-1}$ for $p_F\xi_0=1,2$ and $4$.
Also contributions from scattering states for each $p_F\xi_0$ are plotted.}
 \label{xi}
 \end{wrapfigure}

As mentioned in \ref{sec:op}, the core structure depends on the bound states.
As defined by Kramer and Pesch,\cite{K-P}
we calculate the core radius $\xi_1$ as
\begin{subeqnarray}
\frac{1}{\xi_1}&=&\lim_{r\rightarrow 0}\frac{\Delta(r)}{r\Delta_0},\\
     &=& \frac{g}{\Delta_0}\sum_{E_{n0}\leq E_c}\sum_{j_1j_2}u_{n0j_1}
v_{n0j_2}
\left[1-2f\left(E_{n0}\right)\right]\phi_{0j_1}(0)
\frac{d\phi_{1j_2}}{dr}(0).
\end{subeqnarray}
We plot total $\xi_1^{-1}$ and
the contribution from the scattering states to $\xi_1^{-1}$ in Fig.~\ref{xi}.

The temperature dependence of $\xi_1$ is almost linear in the intermediate
temperature region, according to Kramer and Pesch.~\cite{K-P}
For $p_F\xi_0=4$, there is a substantial decrease of the core radius
at low temperature because of empty bound states.
Although for smaller $p_F\xi_0$ there is a bound state contribution,
the peak position for its contribution to the order parameter is nearly at the
edge of the core.
Therefore the core radius is not strongly affected.
But for large $p_F\xi_0$, the peak of the contribution of the bound states
is well inside the core, so the core radius shrinks significantly.

Our result exhibits saturations of $\xi_1^{-1}$
at $T/T_c\approx 0.1$ for $p_F\xi_0=4$,
at $T/T_c\approx 0.2$ for $p_F\xi_0=2$ and at $T/T_c\approx 0.3$
for $p_F\xi_0=1$.
This saturaion comes from the disappearance of the occupied bound states at low
temperature.
In Fig.~3 in Ref.~\citen{hayasi}, $\xi_1$ is found to saturate
at $T/T_c\approx 0.1$ for small $p_F\xi_0$.
However, the authors of that work did not impose
the particle number conservation condition, 
and therefore the temperature dependence of $\xi_1$ is different from
that in our result.

\begin{figure}
\epsfxsize=13.5cm
\centerline{\epsfbox{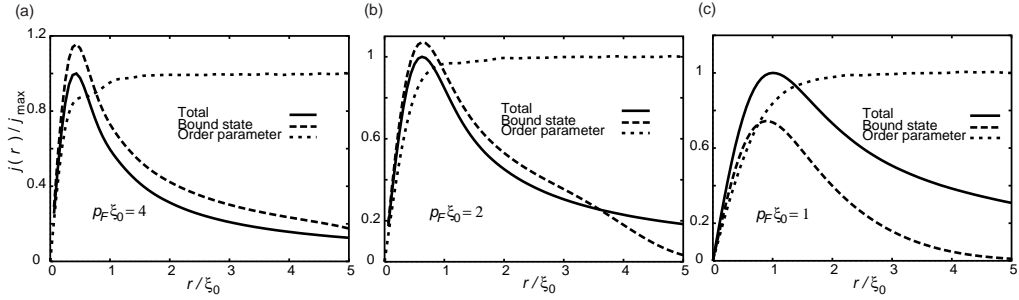}}
\caption{The current density at $T=0.1T_c$ for (a) $p_F\xi_0=4$,
(b) $p_F\xi_0=2$ and (c) $p_F\xi_0=1$. The current density is normalized by
its maximum value. The order parameter is also shown for comparison.}
 \label{current}
 \end{figure}

\subsection{Current density and magnetic field}
The current density is calculated from
\begin{equation}
{\mbf j}({\mbf r})=\frac{e}{2m_ei}\sum_{nm}\left\{f(E_{nm})
u_{nm}^*({\mbf r})\nabla u_{nm}({\mbf r})+\left[1-f(E_{nm})\right]
v_{nm}({\mbf r})\nabla v_{nm}^*({\mbf r}) - \mbox{h.c.}\right\}.
\end{equation}
There exists rotational symmetry, so the current has only a $\theta$ component:
\begin{equation}
j_{\theta}(r)=\frac{e}{m_e}\sum_{nm}\left\{f(E_{nm})
\frac{m}{r}\left|u_{nm}(r)\right|^2-
\left[1-f(E_{nm})\right]\frac{m+1}{r}
\left|v_{nm}(r)\right|^2\right\}.
\end{equation}
 \begin{wrapfigure}{l}{6.6cm}
 \epsfxsize=6cm
  \centerline{\epsfbox{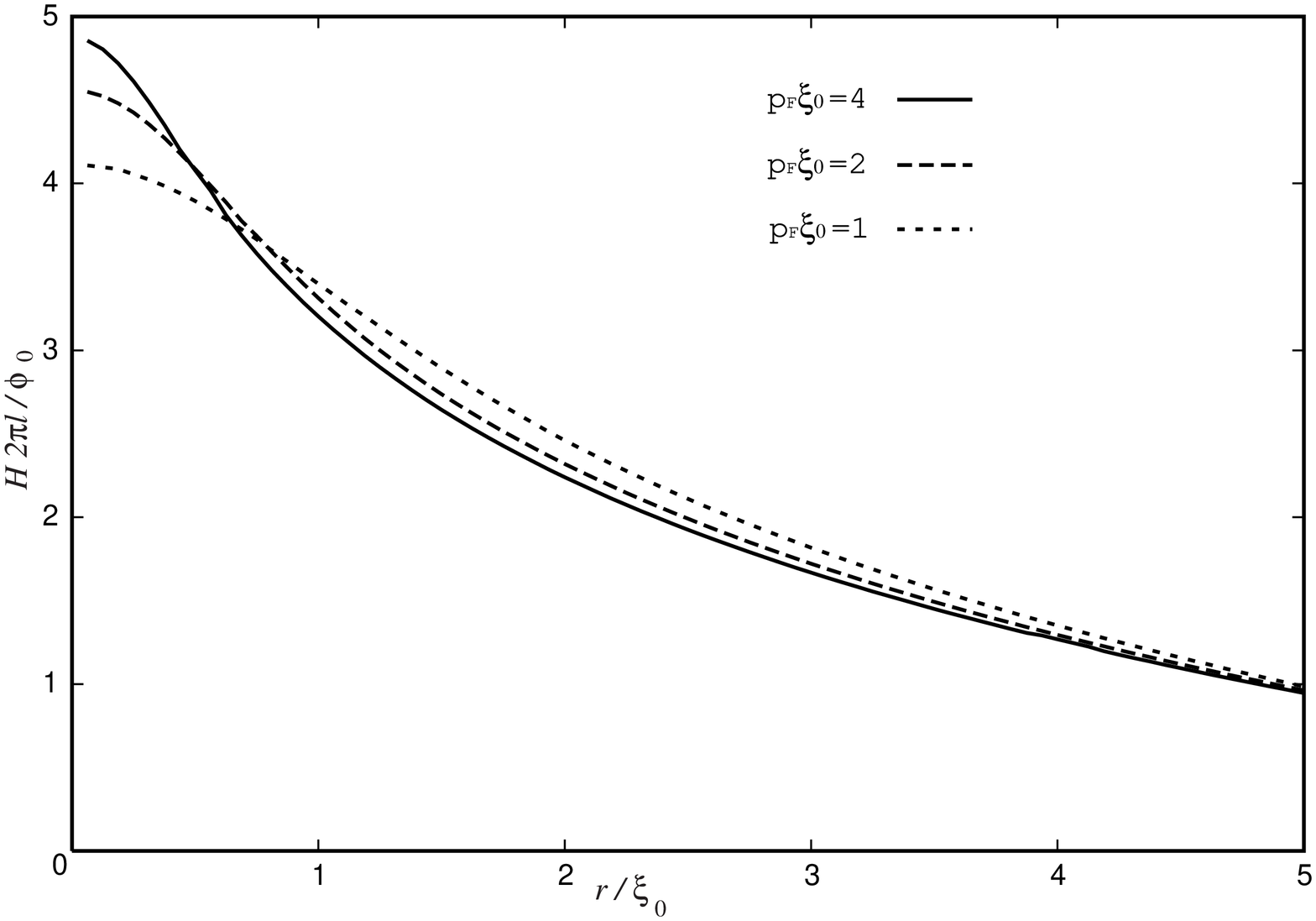}}
 \caption{ The spatial dependence of the magnetic field $H$ at $T=0.1T_c$
  for $p_F\xi_0=1,2$ and $4$. }
 \label{mag}
 \end{wrapfigure}

We show this current density at $T=0.1T_c$ in Fig.~\ref{current}
for each $p_F\xi_0$.
We see that the peak position of the current density is almost the same
as the peak position of
the lowest energy bound state.
Therefore, for a smaller value of $p_F\xi_0$, the peak is located
farther from the vortex core.
Also, for smaller $p_F\xi_0$, the contribution of the scattering states is
the same as that of the bound states,
although for larger $p_F\xi_0$ the contribution of the scattering states is
opposite to that of the bound states.
Therefore there is no simple relation between the core size and
the peak position of the current; that is, they are not exactly the same.

From Maxwell's equations, we calculate the magnetic field parallel to the $z$-axis
as
 \begin{equation}
H_z(r)=\frac{4\pi}{c}\int_r^R j_\theta(r)dr.
\end{equation}
We have normalized this with $\phi_0/2\pi\lambda^2$,
and show in Fig.~\ref{mag},
where $\lambda=\left(m_ec^2/4\pi ne^2\right)^{1/2}$ is the penetration depth
and $\phi_0=hc/2e$ is the flux quantum.
From this figure, we find that for smaller $p_F\xi_0$, the distribution of
the magnetic field is extended farther to the outside of the vortex core.

\subsection{Local density of states}
In order to compare with STM experiments,
we calculate the local density of states
with thermal average,
\begin{equation}
N(r,E)=-\sum_{nm}\{|u_{nm}(r)|^2f'(E_{nm}-E)+
|v_{nm}(r)|^2f'(E_{nm}+E))\}.
\end{equation}

\begin{wrapfigure}{r}{6.6cm}
 \epsfxsize=5.8cm
  \centerline{\epsfbox{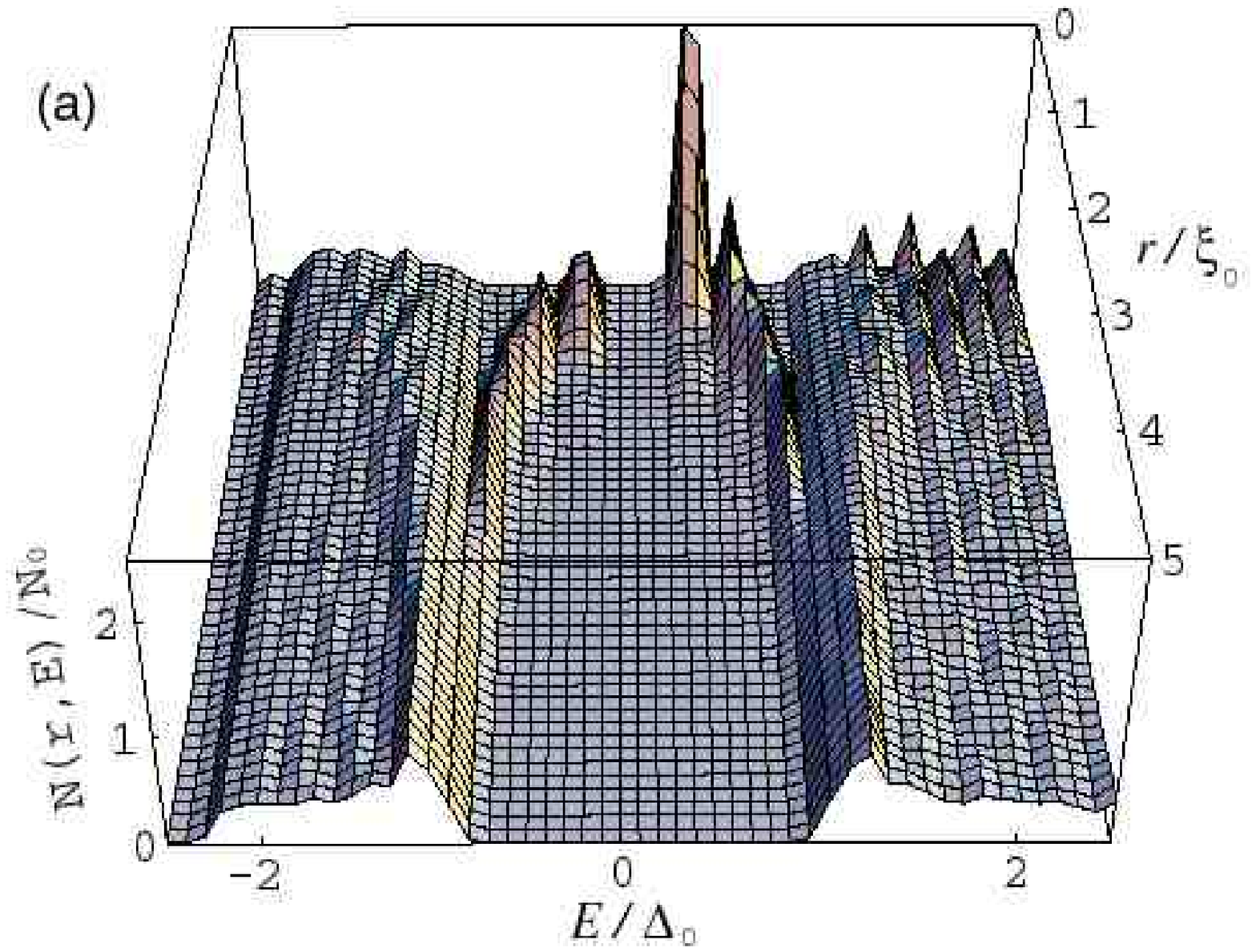}}
  \epsfxsize=5.8cm
  \centerline{\epsfbox{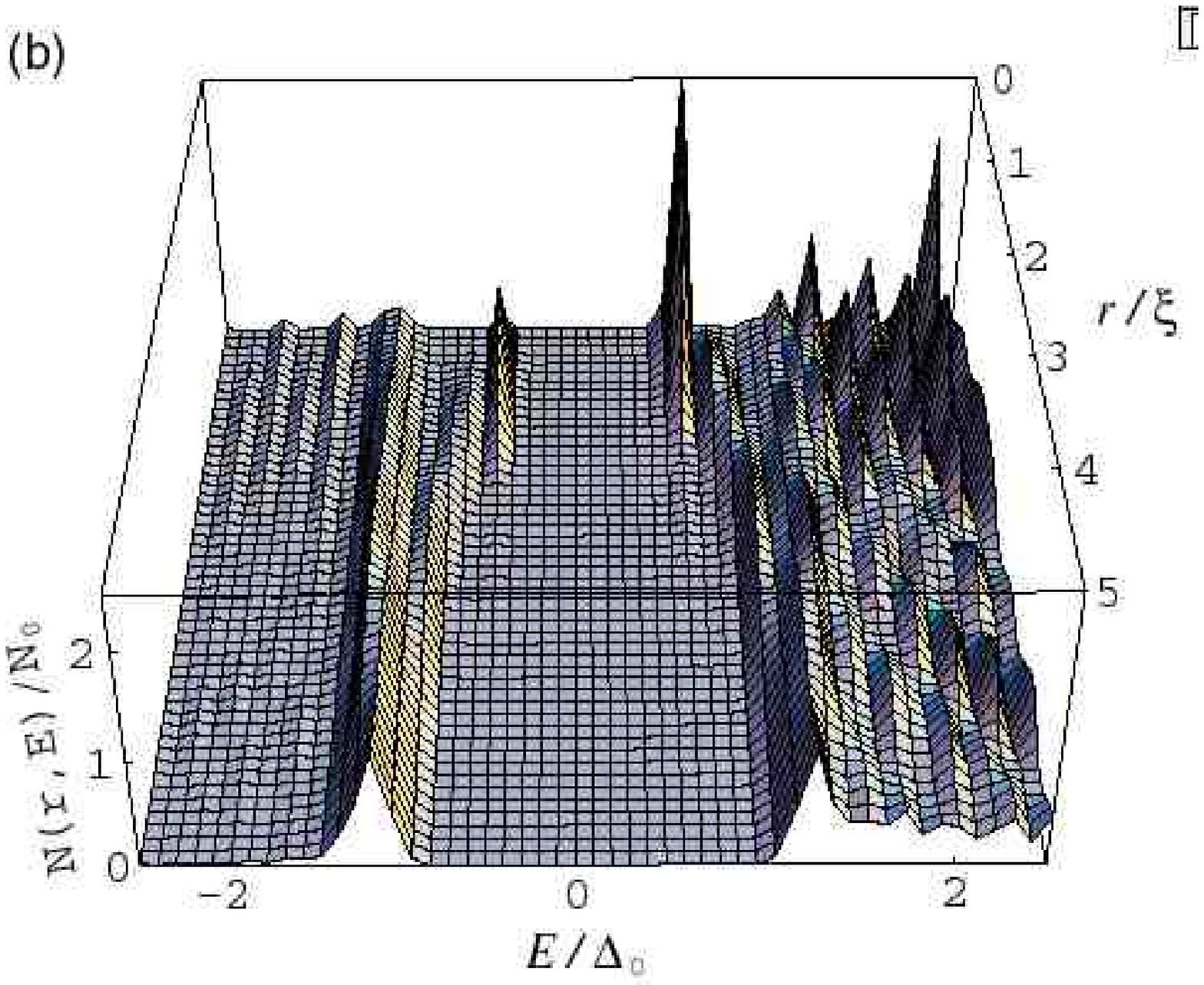}}
  \epsfxsize=5.8cm
  \centerline{\epsfbox{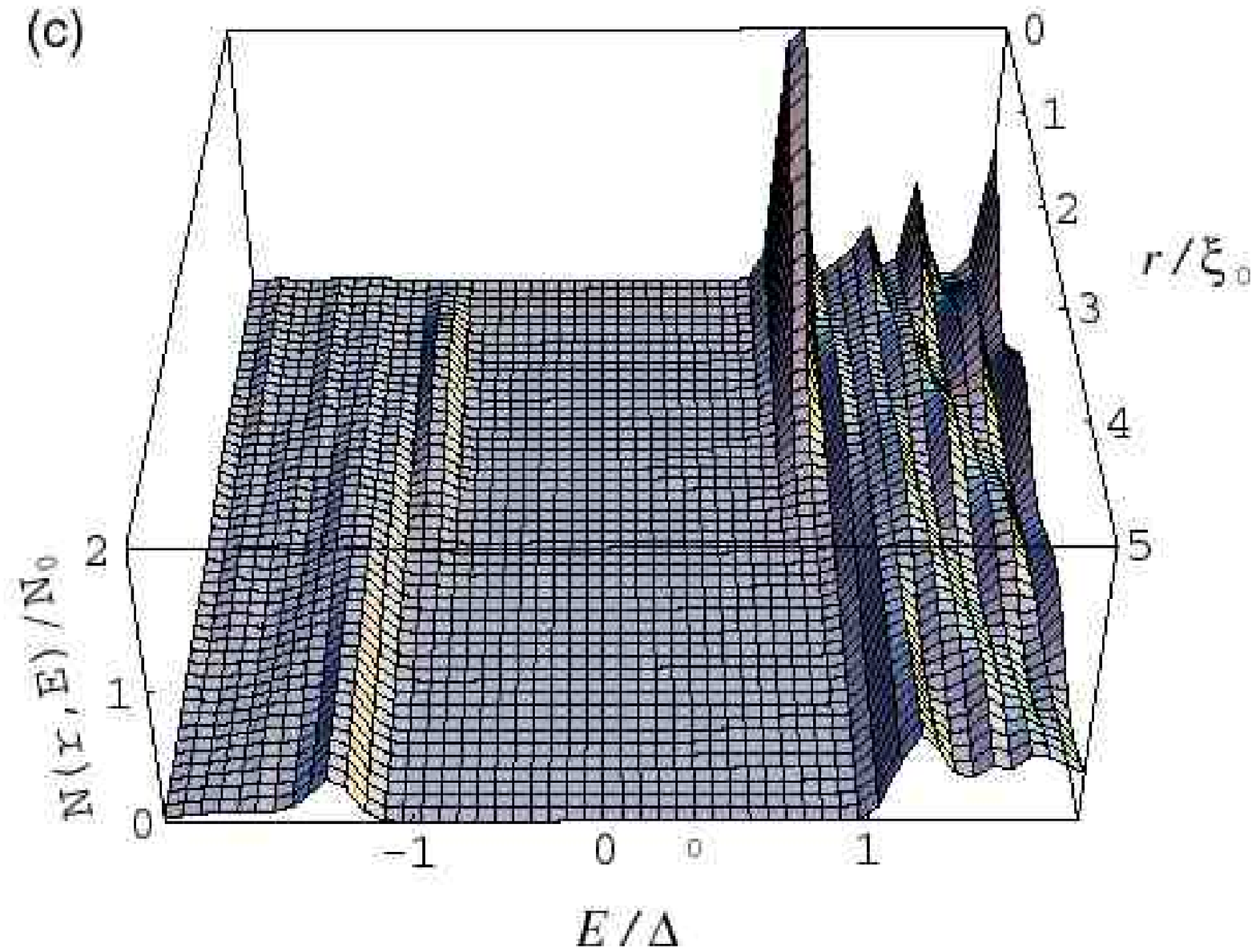}}
\caption{The local density of states $N(r,E)$ at $T/T_c=0.1$ for
(a) $p_F\xi_0=4$, (b) $p_F\xi_0=2$ and (c) $p_F\xi_0=1$.
 $N(r,E)$ is normalized by the density of states of the normal state $N_0$.}
 \label{ldos}
 \end{wrapfigure}

We also take the spatial average with a Gaussian distribution
with standard deviation $0.1\xi_0$.
This is shown in Fig.~\ref{ldos}.
In this figure there is oscillatory behavior for $E/\Delta_0>1.0$, where
$\Delta_0$ is the average
value of the order parameter
outside the vortex core.
This feature results from the finite size of our system, and it is an artifact.
In addition to this, there are several peaks due to the discrete bound states
for energies less than the energy gap ($E<\Delta_0$),
and there is a particle-hole
asymmetry, in contrast to the result of Gygi and Schl\"{u}ter.\cite{gygi-schl}
Wang and MacDonald\cite{wang} demonstrated this asymmetry by solving the lattice
model, and Morita et al.\cite{morita} and Hayasi
et al.\cite{hayasi} pointed it out previously.
At $p_F\xi_0=1$ there is only a peak near the energy gap,
in contrast to the $d$-wave case, where for smaller $p_F\xi_0$ there
are only shoulders inside of the energy gap.\cite{kato}

\section{Conclusion}
In summary, we have solved the Bogoliubov-de Gennes equation with
the number conservation condition
for a single vortex state in $s$-wave superconductors for small $p_F\xi_0$,
and we have obtained the electronic structure around the vortex core.
The number conservation condition strongly affects the quasi-particle
spectrum and the core structure, because the chemical potential varies strongly
with the temperature in the quantum limit ($p_F\xi_0=O\left(1\right)$),
especially for smaller $p_F\xi_0$.
Also, we have elucidated the effect of the discrete bound states
on the order parameter structure,
the local density of states, the current density and the magnetic field.
For smaller $p_F\xi_0$, the contribution of the bound states to
$\Delta\left(r\right)$ decreases, and
the actual shape of $\Delta\left(r\right)$ depends on the peak positions
of the product of the wave functions of bound states.
Also, the discreteness of the bound states is more apparent for smaller
$p_F\xi_0$.

Recently, Sonier et al.\cite{sonier} claimed that they obtained
the core size of the vortex in NbSe$_2$ by measuring the peak position of the
current density.
However, there are a few problems in their procedure.
First, as we have shown, there is apparently no simple relation between
the peak in the current density and the core size.
Second, even if such a relation did exist the relation between these quantities
in the clean limit would be different from that in the dirty limit,
as given by the result in
Ref.~\citen{sonier}.
In particular, the core size in the clean limit is much smaller than that
in the dirty limit with respect to the peak position in the current density.
At least those authors should have analyzed their data in light of
the clean limit calculation by Gygi and Schl\"{u}ter.\cite{gygi-schl}
Third, the muon spin rotation experiment measures the average of the 
spatial distribution of the magnetic field.
Therefore it is not suited to
extract the current distribution.
Rather, it should be used to extract the magnetic-field dependent
magnetic penetration depth, which will provide more insight into the vortex state.

As compared with the result for $d$-wave superconductors,~\cite{kato}
our result exhibits systematic decrease in the number of bound states
with decreasing Fermi momentum.
In the $d$-wave case, there are bound states and extended
states around a vortex.
Therefore detailed study is needed for $d$-wave superconductors.
Such a calculation is now in progress.

\section*{Acknowledgements}
The present work is supported by an NSF under grant number DMR 9531720.
One of us (MK) is supported by the Ministry of Education, Science and Culture
of Japan through the program of exchange researchers.
He is grateful for the hospitality of the Department of Physics and Astronomy of
the University of Southern California.
Also MK appreciates useful discussions with A. Goto and A. Nakanishi.

\end{document}